\documentclass[a4paper]{article}
\usepackage{INTERSPEECH2022}
\title{ESSumm: Extractive Speech Summarization from Untranscribed Meeting}
\name{Jun Wang}
\address{
  University of Maryland, College Park, USA}
\email{junwang@umiacs.umd.edu}

\begin{document}

\maketitle
\begin{abstract}
In this paper, we propose a novel architecture for direct extractive speech-to-speech summarization, ESSumm, which is an unsupervised model without dependence on intermediate transcribed text. Different from previous methods with text presentation, we are aimed at generating a summary directly from speech without transcription. First, a set of smaller speech segments are extracted based on speech signal's acoustic features. For each candidate speech segment, a distance-based summarization confidence score is designed for latent speech representation measure. Specifically, we leverage the off-the-shelf self-supervised convolutional neural network to extract the deep speech features from raw audio. Our approach automatically predicts the optimal sequence of speech segments that capture the key information with a target summary length.  Extensive results on two well-known meeting datasets (AMI and ICSI corpora) show the effectiveness of our direct speech-based method to improve the summarization quality with untranscribed data. We also observe that our unsupervised speech-based method even performs on par with recent transcript-based summarization approaches, where extra speech recognition is required.

\end{abstract}

\section{Introduction}
\label{sec:intro}
Speech is a preferred means for communication among humans\cite{wilpon1994voice}. Automatic speech summarization is a nontrivial and open task in the research of speech understanding. It has wide real-world applications, including broadcast news summarization \cite{maskey2005comparing}, podcast summarization \cite{vartakavi2020podsumm}, clinical conversation summarization \cite{du2019extracting}, and automatic meeting summarization \cite{ murray2008using,  riedhammer2010long}. For example, an automatic meeting summary provides us with an abstract to prepare for an upcoming meeting or review the decisions of previous group \cite{murray2008using}. Given a raw human speech as input, speech summarization is the task of generating a summary presented by speech or text to capture the gist and highlights, without losing significant information.

Our proposed method is aimed at directly generating extractive summary from the original speech in an unsupervised manner, especially on untranscribed meetings. While current literature focus on generating a summary based on transcription, developing an efficient method directly from speech that takes advantage of deep speech features has yet to be explored. Specifically, it is motivated and inspired in three aspects. First, our model is a direct speech summarization method. Most automatic speech summarization frameworks utilize text summarization techniques on top of automatic speech recognition (ASR) output, thus they heavily rely on the availability and quality of ASR engines. Because ASR might not function well for the multiparty spoken dialogue or when applied to languages with limited training resources \cite{flamary2011spoken,tundik2019assessing}, our approach alleviates the problem by directly processing speech signals instead of transcribed text. In this way, our method is applicable to open-domain spontaneous conversations and robust to the possible ASR errors. In addition, ASR system involves different languages' annotation data when applied to different languages, so it is not as general as acoustic characteristics in the real world. Second, our model is an extractive-based method. Summarization techniques are typically classified into two categories, extractive \cite{vartakavi2020podsumm,shang2018unsupervised,zhu2020hierarchical}, and abstractive \cite{mihalcea2004textrank, garg2009clusterrank}. While abstractive summarization can be more concise and flexible, however, extractive summarization can preserve the original format and usually be more fluent \cite{kryscinski2019evaluating}. In addition, the summarized speech from the original utterance is more understandable than the direct transcribed data of speech \cite{rezazadegan2020automatic}. Because of this, our framework is able to naturally and dynamically manipulate the length of generated speech summary, i.e., there is a recent challenge \cite{clifton2020100} where one-minute duration is selected from the Spotify podcast to give the user a sense of what the podcast sounds like. Most importantly, our method helps when the desired summary results are presented by speech instead of text. This opens the door for the applications including the live stream broadcasting \cite{cho2021streamhover}. Last, our approach is fully unsupervised. Compared with the supervised method \cite{zhu2020hierarchical}, no additional annotation data is needed to build the speech summarization model.

To this end,  we design a simple yet efficient speech-to-speech summarization framework, ESSumm, which is aimed at automatically and efficiently summarizing the raw speech input without transcribed data. Our main contributions are three-fold, 
\noindent
\begin{itemize}

 \item To the best of our knowledge, ESSumm is the first architecture  to  explore and incorporate deep  speech  features with  latent semantic analysis  to  the task of meeting speech summarization.
 \item ESSumm is an adaptively extractive speech-to-speech meeting summarization framework in a fully unsupervised manner.
  \item ESSumm could easily concatenate the extracted key speech segments together and produce the short audio summary without additional ASR and speech synthesizer steps. 
\end{itemize}

\begin{figure*}[t]
    \centering
    \includegraphics[width=0.9\linewidth]{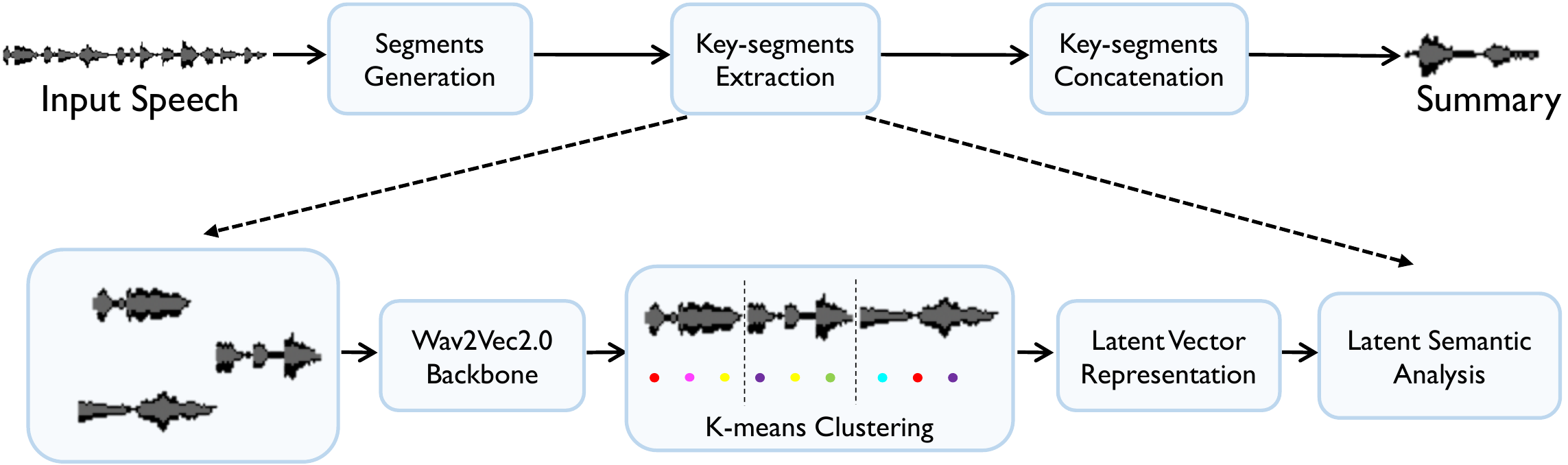}
    \caption{An overview of our ESSumm architecture. It consists of three stages. First, we divide the whole speech input into smaller segments based on the acoustic information. Next, we perform key-segment extraction. Specifically, we extract the deep speech feature representations using pre-trained Wav2Vec2.0 and then project them into high dimensional phoneme probability with k-means clustering algorithm, and score and rank the importance of segments based on Euclidean distance using latent semantic analysis. Finally, we concatenate the key segments together to form the output speech summary with the length constraint. Best viewed in color.}
    \vspace*{-4mm}
    \label{fig:architecture_v1}
\end{figure*}

\section{Related work}
\label{sec:related_work}
\subsection{Acoustic-based speech summarization}
Direct speech summarization approaches have been under-explored in literature. There have been a limited number of prior works that directly deal with speech inputs for automatic speech summarization without transcription \cite{flamary2011spoken,maskey2006summarizing,sert2008combining,zhu2009summarizing,lai2013detecting}. Maskey et al. \cite{maskey2006summarizing} examine Hidden Markov Model on acoustic and prosodic features to select segments for spoken documents, while Flamary et al. \cite{flamary2011spoken} search for the repetitions of recurrent speech patterns. \cite{zhu2009summarizing} summarizes multiple spoken documents by identifying recurring acoustic-based patterns. \cite{sert2008combining} applies the computer vision techniques to  detect possible repetition. Recently, \cite{lai2013detecting} investigates the turn-taking features to examine whether meeting segments contain extractive summary dialogue acts. However, those previous works mainly focus on identifying the recurrent patterns or using the handcrafted features. Instead, our approach summarizes meetings in a distance-based manner from extracted deep speech features.

\subsection{Transcript-based speech summarization}
Existing works \cite{carbonell1998use,mihalcea2004textrank, garg2009clusterrank,zhu2020hierarchical,tixier2017combining, shang2018unsupervised,palaskar2021multimodal} commonly formulate the speech summarization problem to be a text summarization problem in a two-stage process, due to the fact that there have been considerable works and rapid progress in the task of text summarization \cite{nallapati2016abstractive,see2017get,paulus2017deep}. Specifically, they first use ASR engines to generate transcripts from the audio input, which is the word-level information, and then apply the existing mature text summarization methods to produce the summary. Among those transcript-based models, the pioneer maximum marginal relevance (MMR) \cite{carbonell1998use} iteratively selects the most relevant sentence that is similar to the whole document. TextRank \cite{mihalcea2004textrank} is a graph-based keywords extraction algorithm where each keyword is represented by a node in the graph. ClusterRank \cite{garg2009clusterrank} extends TextRank \cite{mihalcea2004textrank} by including measures for noise and redundancy. HMNet \cite{zhu2020hierarchical} implements an encoder-decoder transformer-based network. \cite{tixier2017combining} designs an objective function that is optimized by the greedy algorithm for the extractive summarization in the meeting domain. \cite{shang2018unsupervised} employs a graph-based method for text-based abstractive summarization in an unsupervised manner. \cite{palaskar2021multimodal} extends to model the concept from various modalities. The focus of all above transcribe-and-summarize work, however, is primarily limited to conduct on error-prone transcripts. In this work, we are aimed at achieving summarization in a speech-to-speech manner without transcription.

\subsection{Representation learning on speech}
Recently, there are several works focusing on utilizing the pre-trained speech representations on different speech-related tasks, including speech recognition \cite{chang2021exploration,baevski2020wav2vec}
and speech enhancement \cite{inaguma2020espnet}. Compared with the traditional handcrafted features, including pause, duration, fundamental frequency (F0), and Mel Frequency Cepstral Coefficients (MFCCs), a variety of pre-trained representations obtain significant improvements in the task of speech recognition \cite{chang2021exploration}. Specifically, the recent Wav2Vec2.0 \cite{baevski2020wav2vec} is a transformer-based speech framework that is trained by predicting speech units for masked parts of the audio. Built on the deep speech representations from Wav2Vec2.0 \cite{baevski2020wav2vec}, our summarization framework captures rich relationships among speech segments and produces an effective summary.

\section{Our approach}
\label{sec:Our Approach}
In this work, we focus on the task of extractive summarization. ESSumm takes raw speech as input and generates a summary that covers the most important information. Our goal is to perform speech summarization and generate an extractive summary without transcribed text, where we are able to preserve attributes such as the speaker’s voice, presentation style, type of humor, and production quality. There are mainly two challenges in our task. First, all important key segments should be covered and included in the output summary. Second, the generated summary should be sequential and organized. 

As illustrated in Figure \ref{fig:architecture_v1}, there are three main steps in the proposed architecture, including segments generation, key-segments extraction, and key-segments concatenation. To deal with the first challenge, ESSumm builds on the recent work to extract deep speech feature representations. Specifically, we extract the deep speech feature representations using the pre-trained Wav2Vec2.0 \cite{baevski2020wav2vec}, and then project them into high dimensional phoneme probability with k-means clustering algorithm, and score and rank the importance of segments based on Euclidean distance using latent semantic analysis. To solve the second challenge, we concatenate the key segments together to form the output speech summary with the length constraint.

\subsection{Segments generation}
Speech audio is a continuous signal that captures many aspects of the recording with no clear segmentation into words or other units \cite{baevski2020wav2vec}. Given the raw input speech, most existing summarization works tend to generate the speech segments based on the ASR transcripts, essentially doing text preprocessing to get sentence-level segmentation \cite{vartakavi2020podsumm,zhu2020hierarchical, shang2018unsupervised,palaskar2021multimodal}. Alternatively, we divide the whole input speech into a number of smaller segments based on the acoustic information of silence regions following \cite{lai2013detecting}, where they use ‘spurt’ that separates input speech by at least 500ms silence. In this way, we extract the segments based on the  speech silence gap and don't rely on the availability and quality of ASR engines, which commonly is not optimal under the adverse acoustic circumstances \cite{tundik2019assessing}.

\subsection{Key-segments extraction}
On top of individual speech segments from the first stage, we perform segment scoring and segment selection. Naively, we are able to detect the key segments based on turn-taking information, for example, the absolute difference of average pitch for each speech segment \cite{lai2013detecting}. Alternatively, we propose to leverage pre-trained Wav2Vec2.0 \cite{baevski2020wav2vec} to extract the deep speech-related features for speech segment. Then we use latent semantic analysis based on the metric distance of speech representation to score and rank the importance of candidate segments. 

Motivated by the observation \cite{baevski2020wav2vec} that Wav2Vec2.0 facilitates learning of high-level contextualized representations and has shown its potential in improving the task of speech recognition, we employ the pre-trained Wav2Vec2.0 \cite{baevski2020wav2vec} to extract the deep speech features for each segment. Each speech segment is encoded by the deep speech-related feature representation instead of the hand-crafted acoustic features. In our case of speech summarization, the pre-trained Wav2Vec2.0 model enables us to encode the powerful latent speech representation. 

On the other hand, we leverage latent semantic analysis in this work. First, the speech representation is projected to a high-dimensional phoneme probability. Heuristically, we apply k-means clustering on the extracted deep speech feature representation to get a sequence of phoneme cluster IDs for each segment. In this way, we use the vector speech model representation to represent each segment with a sequence of cluster IDs. As illustrated in the K-means clustering module in Figure \ref{fig:architecture_v1}, different circles for each speech segment refer to its projected vectors. Specifically, different colors correspond to different cluster IDs. Next, we represent each speech segment based on the term frequency–inverse document frequency (TF-IDF) of phonemes. Inverse document frequency (IDF) \cite{jones1972statistical} is a widely used lexical statistical feature in classic information retrieval models. In our case, the TF-IDF vectors are able to capture the phonemes significance. In detail, we adopt the TF-IDF value to measure redundancy and relevance for each speech segment, with the term frequency (TF) being calculated on the segment level and the inverse document frequency (IDF) calculated on the whole input speech. After we get the vector representation for each segment based on TF-IDF, we exploit the principal component analysis (PCA) to represent the whole input speech and we employ the Euclidean distance to the eigenvectors of the whole audio input as the confidence score. Specifically, the Euclidean distance between TF-IDF vectors of each segment and eigenvectors of the whole speech input is inversely proportional to the confidence scores. 

As the key part of ESSumm, the Wav2Vec2.0 deep speech feature extraction module supports efficient and effective speech feature modeling. Because Wav2Vec2.0 learns the speech units common to several languages, our framework benefits from it and is general to multiple different languages. In addition, compared with other supervised meeting summarization approaches, expensive and time-consuming annotations are not required. To this end, we believe that Wav2Vec2.0-based features combined with latent semantic analysis facilitate and directly predict the relative importance of speech segments. 

\subsection{Key-segments concatenation}
Last, we extract the highest-scoring segments to form an extractive summary of the raw input speech audio. We specify the length of the generated summary by a predefined summary time length or by a predefined word count, where we simply generate all the importance-based order of speech segments, and then calculate the length until the target summary length is reached. Most previous works generate the target summary based on the word count. However, the recent challenge \cite{clifton2020100} is aimed at producing one-minute summary from the original podcast. In our case, since the target length is preferred in terms of time length over word number, so ESSumm naturally meets the requirement without an additional speech synthesizer step.

To this end, the quality of generated speech summary can be improved in two aspects. First, the state-of-the-art self-supervised neural network Wav2Vec2.0 \cite{baevski2020wav2vec} accurately captures the speech unit features from raw input audio and learns powerful speech representation from speech segments. To the best of our knowledge, ESSumm is the first work to directly employ a self-supervised neural network on speech summarization task. Second, ESSumm could easily concatenate the extracted key speech segments together and produce a short audio summary without additional ASR and speech synthesizer steps. 

\section{Experiments}
\label{sec:Experiments}
In this section, we evaluate our proposed method on the AMI corpus \cite{carletta2005ami} and ICSI meeting corpus \cite{janin2003icsi}. There has been a shortage of annotated datasets that are tailored for the speech-to-speech summarization task. We begin by giving a brief overview of the datasets in Table \ref{tab:dataset_stat}. Then, we validate the effectiveness of our proposed method by comparing it with the existing approaches on the widely used ROGUE score metric. 

\begin{table}[ht]
    \centering
\begingroup
\caption{Statistics of meeting summarization datasets}
\vspace{-2mm}
\begin{tabular}{l c c} 
 \hline
 \bf Datasets & \bf AMI & \bf ICSI \\ [0.5ex] 
 \hline
\# Meetings  & 137 &  59 \\
\# Speakers  & 4.0 & 6.3 \\
\# Turns & 535.6 & 819.0 \\ 
 \# Len. of Meet. & 6007.7 & 13317.3 \\
 \# Len. of Sum. & 296.6 & 488.5 \\
\hline
\end{tabular}
\vspace*{-5mm}
\label{tab:dataset_stat}
\endgroup

\end{table}

\subsection{Dataset}
\label{dataset}
\noindent \textbf{AMI meeting corpus} 
The AMI corpus \cite{carletta2005ami} is a multi-modal dataset consisting of 100 hours of meeting recordings which is widely used in recent research on meeting summarization. As demonstrated in Table \ref{tab:dataset_stat}, there are 137 meetings in total, each of which involves 4 speakers who work on designing a remote control given variable information. We adopt the traditional test set of 20 meetings following ~\cite{riedhammer2008packing,riedhammer2010long,shang2018unsupervised}. The average length of summary is 296.6 tokens. There is one abstractive summary for each meeting in the test set. The speech recognition transcripts provided in the dataset show a word error rate (WER) of 36\%.

\noindent \textbf{ICSI meeting corpus}
The ICSI meeting corpus \cite{janin2003icsi} is an audio dataset consisting of 70 hours of meeting recordings.
As shown in Table \ref{tab:dataset_stat}, there are 59 meetings in total, each of which involves 6.3 speakers. Different from AMI, the contents of ICSI meetings are specific to the discussions about research among students. We adopt the traditional test set of 6 meetings following ~\cite{shang2018unsupervised,riedhammer2010long}. There are three abstractive summaries for each meeting in the test set. The speech recognition transcripts provided within the dataset show a WER of 37\%.

\subsection{Implementation details}
\label{sec:implementation}

For the segments generation stage, we instead use the provided corpus's dialogue act scripts for once to align with the existing methods. Here, we only use the  sentence-level  information, including the sentence speech segments and their corresponding start-end time in the input audio. Note that we could use the acoustic information to extract segments as well, however, in this case, we are unable to make a fair comparison with existing approaches.  Similar to \cite{shang2018unsupervised}, we randomly select 15 meetings as a development set from the training set for each corpus in order to perform parameter tuning.

\begin{table}[ht]
    \centering
\begingroup
\caption{ESSumm outperforms existing acoustic-based speech summarization methods for ROUGE on AMI corpus test set compared with abstractive ground truth summary using ASR transcriptions. We are using macro-averaged results for 350 word summaries}
\vspace{-2mm}
\setlength{\tabcolsep}{2.4pt}
\renewcommand{\arraystretch}{1} 
\resizebox{0.6\linewidth}{!}{%

\begin{tabular}{l | c c }
\toprule
 & \multicolumn{2}{c}{AMI} \\
Method & ROUGE-2
& ROUGE-SU4 \\
\hline
Murray et al. \cite{murray2008using} & 0.041 & 0.069 \\
\hline
ESSumm & \bf  0.058 & \bf 0.120 \\
\bottomrule
\end{tabular}
}
\vspace*{-2.4mm}
\label{tab:ami_icsi_results_speech_based}
\endgroup
\end{table}

Specifically, we set the number of phonemes to be 32. We project the deep speech features representations obtained from Wav2Vec2.0 \cite{baevski2020wav2vec} to 32 channels. Heuristically, we use four principal components in terms of PCA for AMI and two principal components for ICSI when we calculate the Euclidean distance. Although ESSumm is preferred with speech summary constrained by a target time length, however, it is flexibly adapted to a predefined word count as long as the corresponding timestamp information is provided. In terms of the target summary length, we follow \cite{tixier2017combining,riedhammer2010long} with different word counts for AMI and ICSI, respectively.

\subsection{Evaluation metrics}
\label{evaluation_metrics}
To be consistent with previous works \cite{murray2008using,carbonell1998use, tixier2017combining,shang2018unsupervised}, we adopt the widely used text summarization evaluation metrics: calculating Recall Oriented Understudy for Gisting Evaluation (ROUGE) \cite{lin2004rouge,ganesan2018rouge}. ROUGE is a score of overlapping between the generated summary and the ground truth summary, i.e., ROUGE-1 measures the overlapping of unigrams, and ROUGE-2 refers to the overlapping of bigrams. We utilize the human abstracts as the reference summaries and employ ROUGE-2.0 package \cite{ganesan2018rouge} to evaluate our models. We report the recall, precision and F-1 measure scores for the summarization system performance. 

\begin{table}[ht]
    \centering
\begingroup
\caption{ESSumm outperforms existing sentence-level transcript-based speech summarization methods for all ROUGE levels F-1 measure scores on AMI and ICSI corpora test sets compared with abstractive ground truth summary using ASR transcriptions. Following \cite{riedhammer2010long}, we are using macro-averaged results for 300 and 500 word summaries}
\vspace{-2mm}
\setlength{\tabcolsep}{2.4pt}
\renewcommand{\arraystretch}{1.2} 
\resizebox{0.98\linewidth}{!}{%

\begin{tabular}{l | c c c | c c c }
\toprule
 & \multicolumn{3}{c|}{AMI} & \multicolumn{3}{c}{ICSI} \\
Method & ROUGE-1 & ROUGE-2
& ROUGE-SU4 & ROUGE-1 & ROUGE-2 & ROUGE-SU4  \\
\hline
MMR \cite{carbonell1998use} & 0.24 & 0.04 & 0.07 & 0.10 & 0.01 & 0.02 \\
Concepts \cite{riedhammer2010long} & 0.30 & 0.05 & 0.08 & 0.16 & 0.01 & 0.03 \\
\hline
ESSumm &  \bf 0.35 &   \bf 0.06 &  \bf 0.12 &  \bf 0.30 & \bf 0.04 & \bf 0.10 \\
\bottomrule
\end{tabular}
}
\vspace*{-4.5mm}
\label{tab:ami_icsi_results_sentence_based}
\endgroup
\end{table}

\begin{table}[ht]
    \centering
\begingroup
\caption{ESSumm achieves competitve results compared with existing word-level transcript-based speech summarization methods for ROUGE-1 on AMI and ICSI corpora test sets compared with abstractive ground truth summary using ASR transcriptions. We are using macro-averaged results for 350 and 450 word summaries following \cite{tixier2017combining}. "R", "P", and "F-1" refer to recall, precision and F-1 measure scores. Note that ClusterRank*, Textrank*, CoreRank* are the reproduced results from \cite{shang2018unsupervised} since we use the same official test set split}

\setlength{\tabcolsep}{2.4pt}
\renewcommand{\arraystretch}{1.2} 
\resizebox{0.98\linewidth}{!}{%

\begin{tabular}{l | c c c | c c c }
\toprule
 & \multicolumn{3}{c|}{AMI ROUGE-1 (\%) } & \multicolumn{3}{c}{ICSI ROUGE-1 (\%)} \\
Method & R & P
& F-1 & R & P & F-1  \\
\hline
SummaRunner \cite{nallapati2017summarunner} & - & - & 30.98 & - & - & 27.60 \\
ClusterRank* \cite{garg2009clusterrank} & 39.36 & 32.53 & 35.14 & 
 32.63 & 24.44 & 27.64  \\

TextRank* ~\cite{mihalcea2004textrank} & 39.55 & 32.60 & 35.25 &  
34.89 & 26.33 & 29.70 \\

CoreRank* \cite{tixier2017combining} &  \bf 41.14 &  \bf 32.93 &  
 \bf 36.13 & 35.22 &  \bf 26.34 &  \bf 29.82  \\
\hline
ESSumm & 40.36 & 31.63 & 34.96 & \bf  35.76 & 25.46 & 29.46 \\
\bottomrule
\end{tabular}
}
\vspace*{-5mm}
\label{tab:ami_icsi_results}
\endgroup
\end{table}

\subsection{Experimental results}
\label{results}
Different experimental setups and evaluation criteria, including generated summary type, generated summary length, and reference type, make it challenging to compare previous works \cite{riedhammer2008packing}. Towards it, we manually divide the existing methods into three categories, acoustic-based, sentence-level transcript-based, and word-level transcript-based extractive summarization. Among the latter two, it is nontrivial for the word-level method to produce the summarized speech from the original utterance, while it is practical for the sentence-level approaches as long as the sentence-timestamp pair information is provided.

As demonstrated in Table~\ref{tab:ami_icsi_results_speech_based}, ESSumm outperforms the existing acoustic-based extractive summarization method for both ROUGE-2 and ROUGE-SU4 levels on the AMI corpus test set compared with abstractive ground truth summary using ASR transcriptions. Our results demonstrate that by leveraging state-of-the-art transformer-based acoustic model and latent semantic analysis techniques, a realistic speech summary can be generated to enhance the summarization quality without transcribed data in the real world. Shown in Table~\ref{tab:ami_icsi_results_sentence_based}, ESSumm outperforms existing sentence-level transcript-based extractive summarization methods for all ROUGE levels on AMI and ICSI corpora test sets compared with ground truth abstractive summary using ASR transcriptions. We are using macro-averaged results for 300 and 500-word summaries following \cite{riedhammer2010long}, respectively.

In addition, we compare our framework with the existing word-level transcript-based summarization methods on both AMI and ICSI datasets.  Traditional extractive summarization methods based on selecting and reordering salient words tend to produce summaries that are not natural and incoherent \cite{li2019keep}. As can be seen from Table \ref{tab:ami_icsi_results}, despite its simplicity, our approach works comparably well with previous extractive summarization methods on both AMI and ICSI datasets. Note that although our results are partially constrained by the audio segments specified by the corpus's dialogue act scripts, our method is competitive with the salient word-level methods. 

\section{Conclusions}
\label{sec:conclusion}
In this work, we present a simple yet efficient framework for extractive speech-to-speech summarization, which employs latent speech representation in an unsupervised manner that reduces the requirement of
manual annotations. Unlike the existing work, we can directly generate the summarized speech without the extra speech synthesizer step. Without bells and whistles, experimental results show that the proposed speech-based method is effective for the task of extractive speech summarization in the AMI and ICSI corpora.

\section{Acknowledgements}
The author would like to thank Kishore Prahallad for valuable discussions.

\clearpage
\bibliographystyle{IEEEtran}

\bibliography{mybib}

\begin{thebibliography}{10}
\providecommand{\url}[1]{#1}
\csname url@samestyle\endcsname
\providecommand{\newblock}{\relax}
\providecommand{\bibinfo}[2]{#2}
\providecommand{\BIBentrySTDinterwordspacing}{\spaceskip=0pt\relax}
\providecommand{\BIBentryALTinterwordstretchfactor}{4}
\providecommand{\BIBentryALTinterwordspacing}{\spaceskip=\fontdimen2\font plus
\BIBentryALTinterwordstretchfactor\fontdimen3\font minus
  \fontdimen4\font\relax}
\providecommand{\BIBforeignlanguage}[2]{{%
\expandafter\ifx\csname l@#1\endcsname\relax
\typeout{** WARNING: IEEEtran.bst: No hyphenation pattern has been}%
\typeout{** loaded for the language `#1'. Using the pattern for}%
\typeout{** the default language instead.}%
\else
\language=\csname l@#1\endcsname
\fi
#2}}
\providecommand{\BIBdecl}{\relax}
\BIBdecl

\bibitem{wilpon1994voice}
J.~G. Wilpon, D.~B. Roe \emph{et~al.}, \emph{Voice communication between humans
  and machines}.\hskip 1em plus 0.5em minus 0.4em\relax National Academies
  Press, 1994.

\bibitem{maskey2005comparing}
S.~Maskey and J.~Hirschberg, ``Comparing lexical, acoustic/prosodic, structural
  and discourse features for speech summarization,'' in \emph{Ninth European
  Conference on Speech Communication and Technology}, 2005.

\bibitem{vartakavi2020podsumm}
A.~Vartakavi and A.~Garg, ``Podsumm--podcast audio summarization,'' \emph{arXiv
  preprint arXiv:2009.10315}, 2020.

\bibitem{du2019extracting}
N.~Du, K.~Chen, A.~Kannan, L.~Tran, Y.~Chen, and I.~Shafran, ``Extracting
  symptoms and their status from clinical conversations,'' \emph{arXiv preprint
  arXiv:1906.02239}, 2019.

\bibitem{murray2008using}
G.~Murray, ``Using speech-specific characteristics for automatic speech
  summarization,'' \emph{Diss. Citeseer}, 2008.

\bibitem{riedhammer2010long}
K.~Riedhammer, B.~Favre, and D.~Hakkani-T{\"u}r, ``Long story short--global
  unsupervised models for keyphrase based meeting summarization,'' \emph{Speech
  Communication}, vol.~52, no.~10, pp. 801--815, 2010.

\bibitem{flamary2011spoken}
R.~Flamary, X.~Anguera, and N.~Oliver, ``Spoken wordcloud: Clustering recurrent
  patterns in speech,'' in \emph{2011 9th International Workshop on
  Content-Based Multimedia Indexing (CBMI)}.\hskip 1em plus 0.5em minus
  0.4em\relax IEEE, 2011, pp. 133--138.

\bibitem{tundik2019assessing}
M.~A. T{\"u}ndik, V.~Kasz{\'a}s, and G.~Szasz{\'a}k, ``Assessing the semantic
  space bias caused by asr error propagation and its effect on spoken document
  summarization.'' in \emph{INTERSPEECH}, 2019, pp. 1333--1337.

\bibitem{shang2018unsupervised}
G.~Shang, W.~Ding, Z.~Zhang, A.~J.-P. Tixier, P.~Meladianos, M.~Vazirgiannis,
  and J.-P. Lorr{\'e}, ``Unsupervised abstractive meeting summarization with
  multi-sentence compression and budgeted submodular maximization,''
  \emph{arXiv preprint arXiv:1805.05271}, 2018.

\bibitem{zhu2020hierarchical}
C.~Zhu, R.~Xu, M.~Zeng, and X.~Huang, ``A hierarchical network for abstractive
  meeting summarization with cross-domain pretraining,'' \emph{arXiv preprint
  arXiv:2004.02016}, 2020.

\bibitem{mihalcea2004textrank}
R.~Mihalcea and P.~Tarau, ``Textrank: Bringing order into text,'' in
  \emph{Proceedings of the 2004 conference on empirical methods in natural
  language processing}, 2004, pp. 404--411.

\bibitem{garg2009clusterrank}
N.~Garg, B.~Favre, K.~Reidhammer, and D.~Hakkani~T{\"u}r, ``Clusterrank: a
  graph based method for meeting summarization,'' Idiap, Tech. Rep., 2009.

\bibitem{kryscinski2019evaluating}
W.~Kry{\'s}ci{\'n}ski, B.~McCann, C.~Xiong, and R.~Socher, ``Evaluating the
  factual consistency of abstractive text summarization,'' \emph{arXiv preprint
  arXiv:1910.12840}, 2019.

\bibitem{rezazadegan2020automatic}
D.~Rezazadegan, S.~Berkovsky, J.~C. Quiroz, A.~B. Kocaballi, Y.~Wang,
  L.~Laranjo, and E.~Coiera, ``Automatic speech summarisation: A scoping
  review,'' \emph{arXiv preprint arXiv:2008.11897}, 2020.

\bibitem{clifton2020100}
A.~Clifton, S.~Reddy, Y.~Yu, A.~Pappu, R.~Rezapour, H.~Bonab, M.~Eskevich,
  G.~Jones, J.~Karlgren, B.~Carterette \emph{et~al.}, ``100,000 podcasts: A
  spoken english document corpus,'' in \emph{Proceedings of the 28th
  International Conference on Computational Linguistics}, 2020, pp. 5903--5917.

\bibitem{cho2021streamhover}
S.~Cho, F.~Dernoncourt, T.~Ganter, T.~Bui, N.~Lipka, W.~Chang, H.~Jin,
  J.~Brandt, H.~Foroosh, and F.~Liu, ``Streamhover: Livestream transcript
  summarization and annotation,'' \emph{arXiv preprint arXiv:2109.05160}, 2021.

\bibitem{maskey2006summarizing}
S.~Maskey and J.~Hirschberg, ``Summarizing speech without text using hidden
  markov models,'' in \emph{Proceedings of the Human Language Technology
  Conference of the NAACL, Companion Volume: Short Papers}, 2006, pp. 89--92.

\bibitem{sert2008combining}
M.~Sert, B.~Baykal, and A.~Yazici, ``Combining structural analysis and computer
  vision techniques for automatic speech summarization,'' in \emph{2008 Tenth
  IEEE International Symposium on Multimedia}.\hskip 1em plus 0.5em minus
  0.4em\relax IEEE, 2008, pp. 515--520.

\bibitem{zhu2009summarizing}
X.~Zhu, G.~Penn, and F.~Rudzicz, ``Summarizing multiple spoken documents:
  finding evidence from untranscribed audio,'' in \emph{Proceedings of the
  Joint Conference of the 47th Annual Meeting of the ACL and the 4th
  International Joint Conference on Natural Language Processing of the AFNLP},
  2009, pp. 549--557.

\bibitem{lai2013detecting}
C.~Lai, J.~Carletta, S.~Renals, K.~Evanini, and K.~Zechner, ``Detecting
  summarization hot spots in meetings using group level involvement and
  turn-taking features.'' in \emph{INTERSPEECH}, 2013, pp. 2723--2727.

\bibitem{carbonell1998use}
J.~Carbonell and J.~Goldstein, ``The use of mmr, diversity-based reranking for
  reordering documents and producing summaries,'' in \emph{Proceedings of the
  21st annual international ACM SIGIR conference on Research and development in
  information retrieval}, 1998, pp. 335--336.

\bibitem{tixier2017combining}
A.~Tixier, P.~Meladianos, and M.~Vazirgiannis, ``Combining graph degeneracy and
  submodularity for unsupervised extractive summarization,'' in
  \emph{Proceedings of the workshop on new frontiers in summarization}, 2017,
  pp. 48--58.

\bibitem{palaskar2021multimodal}
S.~Palaskar, R.~Salakhutdinov, A.~W. Black, and F.~Metze, ``Multimodal speech
  summarization through semantic concept learning,'' \emph{Proc. Interspeech
  2021}, pp. 791--795, 2021.

\bibitem{nallapati2016abstractive}
R.~Nallapati, B.~Zhou, C.~Gulcehre, B.~Xiang \emph{et~al.}, ``Abstractive text
  summarization using sequence-to-sequence rnns and beyond,'' \emph{arXiv
  preprint arXiv:1602.06023}, 2016.

\bibitem{see2017get}
A.~See, P.~J. Liu, and C.~D. Manning, ``Get to the point: Summarization with
  pointer-generator networks,'' \emph{arXiv preprint arXiv:1704.04368}, 2017.

\bibitem{paulus2017deep}
R.~Paulus, C.~Xiong, and R.~Socher, ``A deep reinforced model for abstractive
  summarization,'' \emph{arXiv preprint arXiv:1705.04304}, 2017.

\bibitem{chang2021exploration}
X.~Chang, T.~Maekaku, P.~Guo, J.~Shi, Y.-J. Lu, A.~S. Subramanian, T.~Wang,
  S.-w. Yang, Y.~Tsao, H.-y. Lee \emph{et~al.}, ``An exploration of
  self-supervised pretrained representations for end-to-end speech
  recognition,'' \emph{arXiv preprint arXiv:2110.04590}, 2021.

\bibitem{baevski2020wav2vec}
A.~Baevski, H.~Zhou, A.~Mohamed, and M.~Auli, ``wav2vec 2.0: A framework for
  self-supervised learning of speech representations,'' \emph{arXiv preprint
  arXiv:2006.11477}, 2020.

\bibitem{inaguma2020espnet}
H.~Inaguma, S.~Kiyono, K.~Duh, S.~Karita, N.~E.~Y. Soplin, T.~Hayashi, and
  S.~Watanabe, ``Espnet-st: All-in-one speech translation toolkit,''
  \emph{arXiv preprint arXiv:2004.10234}, 2020.

\bibitem{jones1972statistical}
K.~S. Jones, ``A statistical interpretation of term specificity and its
  application in retrieval,'' \emph{Journal of documentation}, 1972.

\bibitem{carletta2005ami}
J.~Carletta, S.~Ashby, S.~Bourban, M.~Flynn, M.~Guillemot, T.~Hain, J.~Kadlec,
  V.~Karaiskos, W.~Kraaij, M.~Kronenthal \emph{et~al.}, ``The ami meeting
  corpus: A pre-announcement,'' in \emph{International workshop on machine
  learning for multimodal interaction}.\hskip 1em plus 0.5em minus 0.4em\relax
  Springer, 2005, pp. 28--39.

\bibitem{janin2003icsi}
A.~Janin, D.~Baron, J.~Edwards, D.~Ellis, D.~Gelbart, N.~Morgan, B.~Peskin,
  T.~Pfau, E.~Shriberg, A.~Stolcke \emph{et~al.}, ``The icsi meeting corpus,''
  in \emph{2003 IEEE International Conference on Acoustics, Speech, and Signal
  Processing, 2003. Proceedings.(ICASSP'03).}, vol.~1.\hskip 1em plus 0.5em
  minus 0.4em\relax IEEE, 2003, pp. I--I.

\bibitem{riedhammer2008packing}
K.~Riedhammer, D.~Gillick, B.~Favre, and D.~Hakkani-T{\"u}r, ``Packing the
  meeting summarization knapsack,'' in \emph{Ninth Annual Conference of the
  International Speech Communication Association}, 2008.

\bibitem{lin2004rouge}
C.-Y. Lin, ``Rouge: A package for automatic evaluation of summaries,'' in
  \emph{Text summarization branches out}, 2004, pp. 74--81.

\bibitem{ganesan2018rouge}
K.~Ganesan, ``Rouge 2.0: Updated and improved measures for evaluation of
  summarization tasks,'' \emph{arXiv preprint arXiv:1803.01937}, 2018.

\bibitem{nallapati2017summarunner}
R.~Nallapati, F.~Zhai, and B.~Zhou, ``Summarunner: A recurrent neural network
  based sequence model for extractive summarization of documents,'' in
  \emph{Thirty-First AAAI Conference on Artificial Intelligence}, 2017.

\bibitem{li2019keep}
M.~Li, L.~Zhang, R.~J. Radke, and H.~Ji, ``Keep meeting summaries on topic:
  Abstractive multi-modal meeting summarization,'' in \emph{57th Conference of
  the Association for Computational Linguistics}, 2019.

\end{thebibliography}

\end{document}